\documentstyle{article}

\if@twoside
\else
\fi
\advance\textheight by \topskip

\makeatletter
\def\@oddhead{\hfill\verb+aipproc2.tex+}
\let\@evenhead\@oddhead
\def\se{\vskip3pt plus1pt minus1pt\setbox0=\hbox to\hsize\bgroup\hss
        \vrule width.5pt
        \vbox\bgroup \hrule width \hsize height.5pt
        \vskip3pt\hbox to\hsize\bgroup\hss\vbox\bgroup\advance\hsize by-9pt
        \columnwidth\hsize\small}
\def\ee{\par\egroup\hss\egroup\vskip3pt\hrule width\hsize height.5pt\egroup
        \vrule width.5pt\hss\egroup
        \box0 \vskip3pt plus1pt minus1pt}
\def\latexe{\LaTeX\kern.15em 2${}_{\textstyle\varepsilon}$}

\makeatother

\begin{document}
\sloppy		

\begin{titlepage}
\vspace*{0pt plus1fill}
\begin{center}
\LARGE\bf
Model Building\\[1pc]
Paul H. Frampton
\end{center}
\vspace*{0pt plus1fill}
\vspace*{0pt plus1fill}
\hbox to\hsize{\hfil
\vbox{\offinterlineskip\hrule height 2pt\vskip4pt
\Large\sf\setbox0=\hbox{May 11, 1997. Vanderbilt University.}\hbox to\wd0{\hfil Talk at Frontiers of Contemporary Physics,
\hfil}
\vskip4pt
\box0
\vskip4pt \hrule height 2pt}%
\hfil}
\end{titlepage}

\newpage

\begin{center}
\LARGE\bf
Abstract
\vspace{2pc}
\end{center}
In this talk I begin with some general discussion
of model building in particle theory, emphasizing 
the need for motivation and testability.
\vspace{1pc}

Three illustrative examples are then described.
The first is the Left-Right model which provides an explanation
for the chirality of quarks and leptons.
The second is the 331-model which offers a first step to
understanding the three generations of quarks and leptons.
Third and last is the SU(15) model which can accommodate
the light leptoquarks possibly seen at HERA.

\newpage

\def\contentsname{\LARGE\bf Contents}

{\large
\tableofcontents
}

\twocolumn

\section{Introduction}

The Standard Model(SM) successfully fits all
reproducible experimental data. The model was built in the 1960s
by Glashow\cite{glashow}, Salam\cite{salam} and Weinberg\cite{weinberg}. 
It was shown to be
renormalizable in 1971 by 't Hooft\cite{hooft}. Its experimental verification
was in excellent shape by 1978 at the Tokyo Rochester Conference\cite{tokyo}.
The $W^{\pm},Z^0$ were discovered in 1983\cite{rubbia}. From 1983 to now there
have been various ambulances to chase, where the data departed from
the SM but further data always rescued the SM. The question is
what to expect as new physics?

\subsection{Fields}

The SM has its own shortcomings and incompleteness 
which motivate model-building
beyond it. A good starting point is to examine the large number of
fields necessary in the SM.

\subsubsection{Gauge Fields}

The gauge sector is based on the group $SU(3)_C \times SU(2)_L \times U(1)_Y$. There are
12 gauge bosons each with two helicity states (we count fields before symmetry breaking)
totalling 24.

\subsubsection{Fermion Fields}

Each of the three generations of quarks and leptons has 15 helicity states. For
the first family there are:
\[
\left( \begin{array}{c} u_{\alpha} \\ d_{\alpha} \end{array} \right)_L \bar{u}^{\alpha}_L
 \bar{d}^{\alpha}_L \left( \begin{array}{c} \nu_{e} \\ e \end{array} \right)_L \bar{e}_L
\]
and similarly in the second and third families:
\[
\left( \begin{array}{c} c_{\alpha} \\ s_{\alpha} \end{array} \right)_L \bar{c}^{\alpha}_L
 \bar{s}^{\alpha}_L \left( \begin{array}{c} \nu_{\mu} \\ \mu \end{array} \right)_L \bar{\mu}_L
\]
\[
\left( \begin{array}{c} t_{\alpha} \\ b_{\alpha} \end{array} \right)_L \bar{t}^{\alpha}_L
 \bar{b}^{\alpha}_L \left( \begin{array}{c} \nu_{\tau} \\ \tau \end{array} \right)_L \bar{\tau}_L
\]

\subsubsection{Scalar Fields}

The minimal scalar sector is one complex doublet of Higgs fields:
\[
\left( \begin{array}{c} \phi^+ \\ \phi^0 \end{array} \right)
\]
with 4 real fields. The total number of fields is thus $24+45+4 = 73$.
In perturbation theory B and L are conserved separately and,
by definition, the
neutrino masses vanish: $m(\nu_i) = 0$.

\subsection{Parameters of the SM}

The model-builders' worksheet.
\begin{itemize}
\item{}Quark masses..................{\bf 6}
\item{}Lepton masses.................{\bf 3}
\item{}Mixing Angles $\theta_i$.............{\bf 3}
\item{}Phase $\delta$............................{\bf 1}
\item{}QCD $\bar{\theta}$.............................{\bf 1}
\item{}Coupling constants..........{\bf 3}
\item{}Higgs sector.....................{\bf 2}
\item{} {\bf TOTAL}........................{\bf 19}
\end{itemize}

There are other fundamental questions to bear in mind:\\

{\it Why three families?}\\

{\it Why $SU(3) \times SU(2) \times U(1)$?}\\

{\it Why these fermion representations?}\\

{\it Elementarity of the Higgs?}\\

\subsection{Model Building}

To build models beyond the standard model is at the same time:
\begin{itemize}
\item{}{\it completely trivial} from the point of view of
a mathematical physicist
- the rules for building renormalizable gauge 
field theories have been known since 1971.
\item{}{\it impossibly difficult} in the absence of
experimental data departing from the SM
 - how can one discriminate between models?
\end{itemize}

\subsubsection{Quality of Model Building}

There are three conditions to judge models by:
\begin{itemize}
\item{}{\bf Motivation}

The model should explain, shed light on, an otherwise unexplained aspect of the SM.

\item{\bf Testability}

At accessible energies either new particles, or rare decays, 
should be predicted to depart clearly from the SM.

\item{}{\bf Aesthetics and Economy}   

Recall Dirac (1928) and his prediction of 
anti-matter. Some physicists, unimpressed by its beauty,
might have discarded the Dirac equation as fatally-flawed 
due to its negative-energy solutions.

\end{itemize}

\subsubsection{Supersymmetry}

The most popular model beyond the Standard Model is
unquestionably supersymmetry.
Its {\it motivation} is to ameliorate, not solve,
the gauge hierarchy problem. It
is clearly {\it testable} by the prediction of a 
large number of sparticles below 1TeV.\\

The models I shall describe have their own
motivation and testability outside of
supersymmetry. This may seem old-fashioned but
it is easier.  After all, any renormalizable gauge theory can
easily be promoted to a globally supersymmetric
model, but the motivation to do so
arises from the mathematical physics end of the spectrum {\it not}
from the physics end.\\

It does seem, on aesthetic grounds, that
supersymmetry is likely to be used by
Nature in Her fundamental theory. The question is 
{\it at what scale} supersymmetry is broken. It
may well be at the Planck scale 
so that supergravity and superstrings
are fine; as for the supersymmetrized SM, 
I would prefer to profess ignorance of physics
above a few PeV, and keep an open mind.\\

Enough proselytising! It is 
time to look at explicit examples.

\section{Left-Right Model}

The {\it motivation} is to understand P violation and secondly
to clarify the role of the (B-L) quantum number.\\

\noindent
{\it Testability} arises from the expectation of non-zero neutrino mass
$m(\nu_i) \neq 0$. Also predicted are $\Delta B = 2$ processes 
like $NN \rightarrow $pions and 
$\Delta L = 2$ processes like neutrinoless double 
beta decay: $(\beta\beta)_{0\nu}$.\\

Assume\cite{PS,MP,SM} the fundamental theory is P invariant. Promote 
the usual electroweak gauge group:
\begin{equation}
SU(2)_L \times U(1)_Y
\end{equation}
to:
\begin{equation}
SU(2)_L \times SU(2)_R \times U(1)_{B-L}  
\end{equation}

The standard relationship:
\begin{equation}
Q = T_{3L} + \frac{Y}{2}
\end{equation}
is replaced by:
\begin{equation}
Q = T_{3L} + T_{3R} + \frac{B - L}{2} \label{Q}
\end{equation}
This is seen by
\begin{equation}
\left( \begin{array}{c} u\\d \end{array} \right)_R
\frac{1}{2}Y = \left( \begin{array}{c} \frac{2}{3} \\ 
\frac{-1}{3} \end{array} \right) =
\left( \begin{array}{c} 
T_{3R} + \frac{B - L}{2} \\
T_{3R} + \frac{B - L}{2} \end{array} \right).
\end{equation}
Similarly for leptons
\begin{equation}
\left( \begin{array}{c} N\\e \end{array} \right)
\frac{1}{2}Y = \left( \begin{array}{c} 0 \\ 
-1 \end{array} \right) =
\left( \begin{array}{c} 
T_{3R} + \frac{B - L}{2} \\
T_{3R} + \frac{B - L}{2} \end{array} \right)
\end{equation}
 
So we now gauge the pre-existing quantity (B - L) rather
than the more arbitrary Y.

The symmetry is broken:
\begin{equation}
SU(2)_L \times SU(2)_R \times U(1)_{B -L} \times P
\end{equation}
\begin{equation}
\stackrel{M_P}{\rightarrow} SU(2)_L \times SU(2)_R \times U(1)_{B-L}
\end{equation}
\begin{equation}
\stackrel{M_{W_R}}{\rightarrow} SU(2)_L \times U(1)_Y
\end{equation}
\begin{equation}
\stackrel{M_{W_L}}{\rightarrow} U(1)_Y
\end{equation}
It is often assumed that $M_P = M_{W_{R}}$. 
For $M_P > \mu > M_{W_{R}}$, $g_{2L}\neq
g_{2R}$ but $W_R$ and $W_L$ remain massless.\\

A minimal Higgs sector contains scalars:
\begin{equation}
\Delta_L(1, 0, +2) ;  \Delta_R(0, 1, +2) ; \phi(\frac{1}{2}, \frac{1}{2}, 0)
\end{equation}
For a range of parameters this gives
a P violating global minimum.\\

For phenomenology, precision tests of the SM require that
$M(W_R), M(Z_R) \geq 500$GeV. Such lower
bounds come from analyses of $\bar{p}p \rightarrow \mu^+\mu^- + X$,
$\mu^- \rightarrow e^-\bar{\nu_e} \nu_{\mu}$, etc.\\

Since (V - A) theory was initially prompted by $m(\nu) = 0$ and
$\gamma_5$ invariance, the presence of (V + A) is naturally linked to 
$m(\nu) \neq 0$ by see-saw formulae like:
\begin{equation}
m(\nu_e) \sim \frac{m_e^2}{M_{W_{R}}}
\end{equation}

\noindent
The right-handed neutrino N is necessary in the left-right
model, just as in the SO(10) GUT.\\

It is natural to expect $\Delta L \neq 0$ Majorana mass 
and $(\beta\beta)_{0\nu}$ proportional to $m_{\nu}$.

From Eq.(\ref{Q}), and using $\Delta Q=0$ while $\Delta T_{3L} = 0$
for $E \gg M_{W_L}$ and given that $\Delta T_{3R} = 1$ we expect transitions
with $|\Delta (B - L)| = 2$.\\

To understand these B violations, it is instructive
to study the partial unification:
\begin{equation}
SU(2)_L \times SU(2)_R \times SU(4)_C     \label{PS}
\end{equation}
in which $SU(4)_C$ contains $SU(3)_C \times U(1)_{B - L}$.
Eq.(\ref{PS}) is a subgroup of SO(10).
By its study we find that $N\bar{N}$ oscillations and
$NN \rightarrow$ pions are predicted.\\

The Left-Right model is the simplest and oldest
extension of the standard model. Amongst its predictions:

\begin{itemize}
\item{}Non-vanishing $m(\nu)$.
\item{}$\Delta B =2, \Delta L =0$ and $\Delta L = 2, \Delta B = 0$ 
processes.
\end{itemize}

\section{331-Model}

The {\it motivation} is to understand better the three
families.\\

\noindent
The {\it testability} is by new accessible particles, most notably
the bilepton gauge boson which can be the doubly charged $Y^{--}$.\\

In the SM, each family separately cancels the trangle anomaly. A
possible reason for three families is that in an extension of the SM
the extended families are anomalous but there is inter-family cancellation.

The three families must enter asymmetrically to set up the +1+1-2
type of anomaly cancellation.

If we assume the first two families are treated symmetrically -
sequentially - then the -2 may be expected to arise from 
$Q(u)/Q(d) = -2$, This {\it is} how it happens.\\

We take the gauge group:
\begin{equation}
SU(3)_c \times SU(3)_L \times U(1)_X
\end{equation}
Hence 331-Model\cite{331}. The standard $U(1)_Y$ is contained in $SU(3)_L$
and $U(1)_X$ while $SU(2)_L$ is a subgroup of $SU(3)_L$. 

The first family of quarks is assigned as:
\begin{equation}
\left( \begin{array}{c} u \\ d \\ D \end{array} \right)_L
\bar{u}_L \bar{d}_L \bar{D}_L
\end{equation}
where the triplet is a ${\bf 3_L}$ of $SU(3)_L$ with $X = -1/3$,
X being the electric charge of the middle component.
The second family of quarks is assigned similarly:
\begin{equation}
\left( \begin{array}{c} c \\ s \\ S \end{array} \right)_L
\bar{c}_L \bar{s}_L \bar{S}_L
\end{equation}

The third family of quarks is assigned differently as:
\begin{equation}
\left( \begin{array}{c} T \\ t \\ c \end{array} \right)_L
\bar{T}_L \bar{t}_L \bar{b}_L
\end{equation}
where the triplet is now ${\bf 3^*_L}$ with $X = +2/3$.

The leptons are treated democratically by assigning:
\begin{equation}
\left( \begin{array}{c} e^+ \\ \nu_e \\e^- \end{array} \right)_L
\left( \begin{array}{c} \mu^+ \\ \nu_{\mu} \\ \mu^- \end{array} \right)_L
\left( \begin{array}{c} \tau^+ \\ \nu_{\tau} \\ \tau^- \end{array} \right)_L
\end{equation}
Each triplet is a ${\bf 3^*_L}$ with $X = 0$.\\

With this arrangement of quarks and leptons, all anomalies cancel. Non-trivial
inter-family cancellations take place for $(3_L)^3$, $(3_L)^2X$, and $X^3$.\\

{\it The number of families must be a multiple of three.}\\

And with six or more families there would be a new replication problem.\\

The Higgs sector is comprised of three triplets and a sextet:
\begin{equation}
\Phi^{\alpha} (X = +1); \Phi^{' \alpha} (X = 0); \Phi^{'' \alpha} (X = -1);
S^{\alpha\beta} (X = 0)
\end{equation}

The scale $U$ at which $SU(3)_L \times U(1)_X \rightarrow SU(2)_L \times U(1)_Y$
has an upper limit for the following reason:\\

The embedding requires that $sin^2\theta < 0.25$. But
at the $Z$ mass, $sin^2\theta > 0.23$ and increases through $0.25$
at a scale $\sim3$TeV. So $U$ must be appreciably below
this to avoid a Landau pole $g_X \rightarrow \infty$.

This implies that $M(Y^{--}, Q) < 1$TeV.\\

The heavy exotic quarks $Q$ can be saught in the usual way.\\

The bilepton can be seen in $e^+e^-$ or hadron colliders. 
Its most striking signature is surely in $e^-e^- \rightarrow 
\mu^-\mu^-$ (NLC) or $\mu^-\mu^- \rightarrow e^-e^-$
(muon collider) where it appears as a sharp peak in the cross-section.
A detailed plot can be found in \cite{331}.

\section{SU(15) Model}

The {\it motivation} is unification without proton decay.\\

\noindent
The {\it testability} is the prediction, possibly recently confirmed,
of weak-scale scalar leptoquarks.\\

In the conventional GUT paradigm, {\it e.g.} $SU(5)$, the couplings
$\alpha_1, \alpha_2, \alpha_3$ unify at $M_{GUT}$ to a single
$\alpha_{GUT}$, hence offering a constraint on the $\alpha_i$.
With $SU(5)$, and likewise the larger unifying groups
$SO(10)$ and $E(6)$, however, {\it proton decay} occurs at leading
order mediated by gauge bosons with definite $(B-L)$ but
indefinite separate $B$ and $L$. \\

$SU(15)$ was proposed in 1990 as a GUT in which gauge mediated
proton decay is absent. It has renewed interest in 1997 because it
predicts light leptoquarks.\\

The symmetry-breaking pattern is:
\begin{equation}
SU(15) \stackrel{M_{GUT}}{\rightarrow} SU(12)_q \times SU(3)_l
\end{equation}
\begin{equation}
\stackrel{M_B}{\rightarrow} SU(6)_L \times SU(6)_R 
\times U(1)_B \times SU(3)_l
\end{equation}
\begin{equation}
\stackrel{M_A}{\rightarrow} SU(3)_C \times SU(2)_L \times U(1)_Y
\end{equation}
\begin{equation}
\stackrel{M_W}{\rightarrow} SU(3)_C \times U(1)_Q
\end{equation}
Once $M_A$ is chosen, $M_B$ and $M_G$ follow from renormalization 
group analysis. It is natural to put $M_A = M_W$ in which case 
all scales are fixed. The families are each put in
one fundamental representation, {\it e.g.}
\begin{equation}
15_L = (u_1, u_2, u_3, d_1, d_2, d_3, \bar{u_1}, \bar{u_2}, 
\bar{u_3}, \bar{d_1}, \bar{d_2}, \bar{d_3}, e^+, \nu_e, e^-)
\end{equation}
Anomaly cancellation is by mirror fermions.\\

The scales $M_B$ and $M_G$ follow from $M_A$ as:

$$\begin{tabular}{||c|c|c||} \hline
$M_A$(GeV) & $M_B$(GeV) & $M_{GUT}$(GeV) \\ 
(INPUT) & & \\  \hline \hline
$250$ & $4.0 \times 10^6$ & $6.0 \times 10^6$	 \\ \hline
$500$ & $5.8 \times 10^6$ & $8.9 \times 10^6$ \\  \hline
$10^3$ & $8.3 \times 10^6$ & $1.3 \times 10^7$ \\ \hline
$2 \times 10^3$ & $1.2 \times 10^7$ & $1.9 \times 10^7$ \\ \hline \hline
\end{tabular}$$

\noindent
The reason gauge bosons have a definite $B$ and $L$ is
because there are only defining representation {\bf 15}s
for the fermions. The {\bf 10} in $SU(5)$ is antisymmetric
and this couples to gauge bosons with indefinite $B$ and $L$.\\

The breaking at $M_W$ involves the SM Higgs doublet which
is contained in an $SU(15)$ {\bf 120}:
\begin{equation}
15 \times 15 = 120_S + 105_A
\end{equation}
Under 3-2-1, this decomposes into:
\begin{equation}
120_S = (3,2)_{+7/3} + (1,2)_{+1} + .......
\end{equation}

\noindent
The two terms exhibited are a candidate scalar leptoquark:
\begin{equation}
\left( \begin{array}{c} e^+u \\ e^+d \end{array} \right)
\end{equation}
and the standard Higgs doublet respectively. Since these are in a common
$SU(15)$ irreducible representation, the leptoquark is predicted
at the weak scale\cite{SU15}.\\

In February 1997, the two major collaborations at HERA, H1\cite{H1} and 
ZEUS\cite{ZEUS}, both announced a small number of events 
consistent with a direct-channel resonance in $e^+p$ with mass $\sim200$GeV. It
is most easily interpreted as involving valence quarks
($u$ or $d$), and to be scalar rather than vector.\\

It is therefore naturally accommodated in the $SU(15)$ model
as the $(3,2)_{+7/3}$ scalar leptoquark.\\

\begin{itemize}
\item{}{\it Several further leptoquark states are predicted.}
\end{itemize}

Predicted leptoquarks in the {\bf 120} of $SU(15)$, at or near the weak scale,
with $(3_C, 2_L, Y)$:

\begin{itemize}
\item{}B=+1/3,L=+1. $(3,3,-2/3)+(3,1,-2/3)$
\item{}B=+1/3,L=-1. $(3,2,+7/3)$   (discovered?)
\item{}B=-1/3,L=-1. $(\bar{3},2,-1/3)+(\bar{3},2,-7/3)$
\item{}B=-1/3,L=-1. $(\bar{3},1,+2/3)+(\bar{3},1,+8/3)$
\end{itemize}

\noindent
The conjugate representation, ${\bf 120^*}$, is 
necessarily also present.\\

\begin{itemize}
\item{}{\it Thus, if the $SU(15)$ interpretation of the HERA data 
is correct, several additional states should be discovered shortly.}
\end{itemize}

\section{Summary}

The examples were intended to illustrate how to {\it motivate}
a model by looking at the parameters of the SM, as well
as the gauge group and matter representations 
{\it e.g.} the three families.\\

Whether a given model will be {\it testable} is
often hard to see in advance and
can only be checked {\it a posteriori}.\\

\begin{itemize}
\item{}Let's hope some challenging models will
be added before LHC is commissioned!
\end{itemize}

\section*{Acknowledgement}

This work was supported in part by the US Department of Energy
under Grant No. DE-FG05-97ER41036.

\end{document}